\documentclass[aps,twocolumn]{revtex4}

\newcommand{\fft}[2]{\frac{#1}{#2}}
\newcommand{\ft}[2]{{\textstyle\frac{#1}{#2}}}

\long\def\symbolfootnote[#1]#2{\begingroup%
\def\thefootnote{\fnsymbol{footnote}}\footnote[#1]{#2}\endgroup}

\begin{document}

\preprint{MCTP-05-94}

\title{Devaluation: a dynamical mechanism for a naturally small
cosmological constant}

\author{Katherine Freese}
\email{ktfreese@umich.edu}

\author{James T.~Liu}
\email{jimliu@umich.edu}

\affiliation{Michigan Center for Theoretical Physics,
University of Michigan,
Ann Arbor, MI 48109-1120}

\author{Douglas Spolyar}
\email{dspolyar@physics.ucsc.edu}

\affiliation{Physics Department, University of California,
Santa Cruz, CA 95060}

\begin{abstract} 
We propose a natural solution to the cosmological constant problem
consistent with the standard cosmology and successful over a broad
range of energies.  This solution is based on the existence of a new
field, the devaluton, with its potential modeled on a tilted
cosine. After inflation, the universe reheats and populates the
devaluton's many minima.  As the universe cools, domain walls form
between different regions.  The domain wall network then evolves and
sweeps away regions of higher vacuum energy in favor of lower energy
ones.  Gravitation itself provides a cutoff at a minimum vacuum
energy, thus leaving the universe with a small cosmological constant
comparable in magnitude to the present day dark energy density.
\end{abstract}

\maketitle


The discrepancy between the observed vacuum energy density of the
universe today and the much larger value expected from the
standard model of particle physics is known as the cosmological
constant problem.  At the same time, however, recent observations of
Type IA Supernovae \cite{riess, perlmutter} indicate
that the universe is accelerating.  One possible explanation of the
``dark energy'' driving this acceleration is that it is due to a
cosmological constant of $(0.003\,\mathrm{eV})^4$; we take this as an
upper limit on the allowed value.  However, this value of the cosmological
constant is unnaturally small in particle physics.

Unnatural cancellations of zero point energies for particles and for
vacuum energies generated at phase transitions are required to one
part in $10^{120}$ so that the sum of all these terms is small enough
to agree with the observations. The hope is sometimes voiced that a
fundamental quantum theory of gravity will resolve this problem, but
such a theory must in fact give rise to a cosmological term (say, at
the Planck scale) which is precisely cancelled by all lower energy
contributions ({\it e.g.}\ at the electroweak scale and below) to one part in
$10^{120}$.  Thus the cosmological constant problem is essentially a
difficulty of physics at very `low' energies, suggesting that its
solution will come instead from new physics manifest at low energies.
One possibility invokes compensating fields whose vacuum energy
dynamically adjusts to cancel the large value described above which
arises from particle physics.

In this letter, we propose a dynamical solution to the cosmological
constant problem (including any contributions generated at phase
transition) involving the post-inflation evolution of a domain
wall network.  What is required is a model with a sufficiently rich
potential which has nearly degenerate minima (separated by barriers
of appropriate height) spanning a range of both negative and positive
values of vacuum energy.  After a conventional stage of inflation
followed by reheating, different patches of the universe fall into
different minima, and a domain wall network is thus formed.

At this stage, the universe contains many different regions with
different vacuum energies separated by domain walls.  However, a
domain wall separating two regions with different vacuum energies
will feel a pressure driving it to move rapidly into the region of
higher energy.  As a result, the domain walls end up sweeping away
regions of high positive energy, thus driving the universe to lower
and lower vacuum energy.

It is of course important to this scenario that the universe
does not end up falling into a state with large regions of negative
vacuum energy.  We will show below that this does not happen, as
gravity differentiates between positive and negative energies.  The
essential feature here is similar to the Coleman and de Luccia
result that tunneling into an anti-de Sitter vacuum is greatly
suppressed in the presence of gravity.  In the present case, however,
this argument involves the evolution of domain walls (as opposed to
instantons) interpolating between different vacuum states.  As a
result, the universe naturally evolves to a final (metastable) state
with the potential sitting at the minimum closest to zero energy.

Given an appropriate potential, this final value of the energy density
could be made on the order of $(10^{-3}\,\mathrm{eV})^4$ in agreement with
observational measurements of the dark energy.  Ideally, one would like
to see this value come out naturally as part of the solution to the
cosmological constant problem.  However, at present we are unable to
claim success in this regard, and simply view this result as imposing
a requirement on the landscape of the minima of the potential.

We denote this process of domain wall evolution driving the universe to
a state of nearly zero vacuum energy by ``devaluation'', the process of
moving towards lower and lower values of the cosmological constant.  As
a concrete realization of this devaluation model, we may consider a
system with two scalar fields, the first an inflaton $\psi$ which
drives inflation and the second the devaluton $\Phi$ which is responsible
for the domain wall network.  Of course, the overall evolution of
the universe is governed by the nature of the coupled potential
$V(\psi,\Phi)$ (which includes the vacuum energy of the standard
particle physics sector).  However, with an appropriate separation of
scales, we may treat the evolution of the universe in two sequential
stages, the first being inflation and the second devaluation.  Thus
we take
\begin{equation}
V(\psi,\Phi)=V_{\rm inf}(\psi)+V_{\rm dev}(\Phi) .
\end{equation}
The energy scales of the inflaton can be anything from Grand Unified
(GUT) scale down to 10 MeV, as long as reheating above nucleosynthesis
is successful.  The devaluton too can have a wide range of energy
scales: $V_{\rm dev}$ can become negative, but its magnitude is
roughly bounded by the energy scale of the inflaton.  In particular,
we absorb any uncontrolled contribution to the vacuum energy into the
inflaton potential $V_{\rm inf}$.

{\it Inflation Field.}
In the inflationary stage, the vacuum energy
dominates the energy density, and the scale factor of the universe
expands superluminally to solve the horizon, flatness and monopole
problems of standard cosmology \cite{guth}.  The inflaton potential
may come from either a tunneling field or a rolling field.  In either
case the important features are sufficient inflation followed by
reheating, with the potential at the end of inflation 
at some positive value $\geq 10\,\mathrm{MeV}$.  It is this remaining amount of
vacuum energy that is removed by the devaluation process.  We
emphasize that the details of the inflaton are {\it not} important for
our devaluation idea; in the next few paragraphs we present an example
of a possible inflaton field.

For a dynamical solution to the cosmological constant problem, we
cannot simply postulate the absolute minimum of the inflaton potential
$V_{\rm inf}$ to be zero.  Thus this requirement that $V_{\rm inf}$
ends up at a sufficiently small positive value at the end of inflation
is not a trivial one.  It turns out, however, that this requirement is
naturally satisfied in the recently proposed chain inflation model
\cite{fs}, which resolves the problems of old inflation while
retaining the idea of tunneling fields dominating the energy density
of the universe.  

In chain inflation, one considers a potential $V_{\rm inf}$ that has
many minima of different vacuum energies (above and below zero) with
barriers separating the different minima.  The field quantum starts out
in a high energy minimum and quantum mechanically tunnels through a series
of barriers from higher to ever lower minima.  The universe inflates by a
fraction $N_{\rm stage} < 1/3$ of an $e$-fold while sitting in each minimum,
adding up to at least 60 $e$-folds from all the tunneling events
($N_{\rm tot}>60$). The number of minima that the field must go through
must be at least 200.

A simple model for the chain inflation potential is a cosine with a linear
tilt
\begin{equation}
\label{eq:chain}
V_{\rm inf}(\psi)=V_0\biggl[1-\cos\fft{\psi}{v_1}\biggr]-\alpha\psi.
\end{equation}

Due to this tilt, any uncontrolled amount of vacuum energy may be
absorbed by a shift in $\psi$.  Of course, to be realistic, this
potential ought to be cut off at an arbitrary energy below zero.
However the details of the cutoff are unimportant as the field never
tunnels completely to the minimum.  While the energy scales of this
potential can be anywhere from the GUT scale down to
$10\,\mathrm{MeV}$, we assume the width, barrier height and tilt of
the potential are all roughly comparable.  For example, if the energy
scale $\sim1\,\mathrm{GeV}$, in the absence of any tuning, there will
typically be a minimum at $V_{\rm inf} \sim \ft12\,\mathrm{GeV}$ and
another at $V_{\rm inf}\sim -\ft12\,\mathrm{GeV}$.  The inflaton then
starts in a high energy minimum and subsequently undergoes a series of
tunneling events until it stops tunneling and gets stuck at the last
positive energy minimum.

The reasons why the universe does not enter a negative energy vacuum are
twofold.  Firstly, Coleman and de Luccia \cite{Coleman:1980aw} studied
tunneling in the thin wall limit and concluded that, for large enough
bubbles, tunneling is suppressed to AdS vacua. They also found that,
if tunneling were to take place, the result would be a negative curvature
universe collapsing to a Big Crunch.  Subsequently,
Banks \cite{Banks:2002nm} generalized these results to claim that decays
into spaces of negative cosmological constant do not occur in consistent
theories of quantum gravity as the requisite instantons for tunneling into
AdS simply cannot exist.

The second argument for why the field gets stuck at the last positive
vacuum, rather than decaying to AdS, is as follows.  At each stage of
tunneling, bubbles are produced of the lower energy vacuum.  Because
$N_{\rm stage}<1/3$, the bubbles percolate at each stage. However, in
the earliest stages of chain inflation, the interiors of the bubbles are
still dS so they cannot thermalize. Only once the field has tunneled many
times and is approaching Minkowski space can the interiors really thermalize.
At this stage one must consider the tunneling in the presence of radiation.
This process has previously been considered by Affleck and de Luccia
\cite{adel}, who found that the presence of particles can either enhance
or reduce the rate of decay of the false vacuum.  By properly adjusting
the coupling, the tunneling rate can be made very slow.  However this
suppression is only relevant in the present case for $V_{\rm inf}$ near
zero since that is where the radiation becomes thermalized.  Hence
this mechanism can be used to allow tunneling at high energies but to
delay tunneling into AdS vacua. We note that rolling fields may 
alternatively drive the inflation, as long as the vacuum at the 
end of inflation reheats to a positive energy value.

{\it Devaluation Field.}  At this stage, the inflaton field has served
its purpose in driving inflation (and reheating in the last few stages
of chain inflation).  After inflation, the universe is radiation
dominated and proceeds through its ordinary evolution, including
primordial nucleosynthesis.  With $V_{\rm inf}$ sitting in a positive
energy minimum $\geq 10\,\mathrm{MeV}$, it is now the r\^ole of the
devaluton to further drive the vacuum energy to near zero.  This is
the most important feature of our model.  To do so, we may take a
periodic devaluton potential of tilted cosine form
\begin{equation}
\label{eq:pot}
V_{\rm dev}(\Phi) = V_0\biggl[1-\cos {N\Phi \over v} \biggr]
 - \eta\cos\biggl[{\Phi \over v} + \gamma \biggr] .
\end{equation}
This potential is bounded and has many minima with large barriers in
between.  The bumps in the potential are present at any temperature, but
only become important once the temperature drops to the value of the
barrier height.
Furthermore, this potential is similar to that of the QCD
axion, where the periodicity in the first term is due to instanton
effects and the tilt in the second term can be produced with a
soft-breaking term (or by coupling to an additional scalar field).
 Such a small tilt, crucial to our model, is then natural as it arises
from soft breaking of a symmetry.  

Defining $\theta = N\Phi/v$, the potential has $N$ minima at $\theta =
2 \pi n$ where $n=0,1,2,\ldots,N-1$ and we take $N$ to be a large
number.  The barrier height can be expressed in terms of the
devaluton mass $m$ as
\begin{equation}
\label{eq:bht}
V_0 = m^2 v^2 /N^2,
\end{equation}
and the wall tension is
\begin{equation}
\label{eq:walltension}
\sigma = 8 m v^2 /N^2.
\end{equation}
The second term tilts the potential and
thereby breaks the degeneracy between the minima, where
the energy difference between minima is roughly 
\begin{equation}
\epsilon
\sim 2 \pi \eta /N ,
\end{equation}
and $\gamma$ is a phase.
Due to the tilt in the potential, the different minima of the
potential have different energies, ranging from $-\eta$ to $+\eta$.
To ensure the possibility of zero vacuum energy, we must take $|\eta|$
to be larger than $V_{\rm inf}$ at the end of chain inflation.  Furthermore,
the energy difference $\epsilon$ must be chosen on the order of
$(10^{-3}\,\mathrm{eV})^4$ to be consistent with present observations.  This
demonstrates that $N$ must be very large, on the order $10^{10}$ or more.

At the end of inflation, but at temperatures much larger than the
devaluton barrier height, the barriers go unnoticed and the universe
populates all values of $\Phi$ (with bias towards lower energies in
the tilt).  However, once the temperature drops down to the value of the
barrier height $V_0$, the barriers become important.  Different patches
of the universe fall into different minima of the devaluton potential
separated by domain walls.  Note that $\Phi$ does not tunnel from one
local minimum to the next as the barrier height between minima is too
large.  However, because the potential is tilted, the different minima
of the potential have different energies, and this provides a bias
driving the domain wall evolution.

This domain wall evolution was previously investigated by
Sikivie \cite{sikivie} in the context of the QCD axion.  In that
context the tilt in the potential was introduced to eliminate excess
domain walls by driving them away.  One way to see this is to note that domain
wall networks with non-degenerate vacuum states mainly evolve due to two
competing effects: a surface tension that acts to straighten the walls
and a volume pressure due to the energy difference between the minima.
The latter force causes the walls to move into the region of higher
energy, thereby making such regions smaller.  This behavior has been
shown to be rather varied and interesting \cite{Gelmini:1988sf}.
Provided the dynamical timescale for this evolution is sufficiently
rapid, the domain walls thus evolve to eliminate regions of higher
energy vacuum in favor of lower energy ones.  In this way, the
universe, which consists of a patchwork of regions with different
values of the devaluton field (sitting in different minima of the
devaluton potential), is driven to a very low value of the
cosmological constant.

As for the timescale, Larsson {\it et al.}\ \cite{larsson} have
studied the evolution of non-degenerate domain wall networks to see
how quickly they disappear. They consider two sources of instability:
first the pressure due to the energy difference between minima, and
second any bias in the original probability distribution that prefers
one vacuum over the other. In our case, there is the obvious bias
incurred by the fact that we are considering a radiation dominated
universe: a Boltzmann distribution favors the lower energy vacuum.
However, near the minimum of the potential the distribution should be
rather flat. As shown in \cite{larsson}, even a tiny statistical bias
between two minima plays an important role in eliminating the less
favored minimum.  They performed numerical simulations to show that,
when the typical domain size has grown to a critical radius $R>R_c
\sim \sigma/\epsilon$, the wall energy density decays exponentially
fast.  The pressure acts very rapidly (faster than on cosmological
timescales) to eliminate the high energy vacua in favor of lower
energy ones. 

Furthermore, the case of adjacent regions of different vacuum energy
has previously been carefully considered by Cveti\v{c} {\it et
  al.}\ \cite{Cvetic1,Cvetic2,Cvetic3}.  They studied the case of a
single domain wall separating two regions of different vacuum energy
which can be either positive energy (dS), zero energy (Minkowski), or
negative energy (AdS).  In their study, there was no radiation in the
system, merely vacuum. They explicitly computed the dynamics of two
regions separated by a domain wall by making an ansatz for the metric
on two sides and connecting them via the Israel matching conditions on
the domain wall boundary.

This devaluation process, whereby domain walls sweep away regions of
high positive energy and drive the universe to vacuum energy near zero
(or to the current value of the dark energy), provides a natural
solution to the cosmological constant problem. We note that, unlike
other proposed solutions to the cosmological constant problem,
reheating is successful in this model. Here radiation was produced at
the end of inflation, leading to the usual radiation and matter
dominated evolution of the universe.  In fact radiation is a critical
element in the devaluation process.  Because the universe is radiation
(or matter) dominated (rather than vacuum dominated), the horizon
grows in time.  Hence neighboring regions which were initially causally
disconnected become causally connected as time goes on.  Thus domain
walls sweep across ever larger regions of the universe, so that larger
and larger patches of the universe are driven to the same low value of
the vacuum energy.  The radiation of course is also a key element in
ending up with a universe that looks like ours which contains matter
and radiation.

{\it Negative Energy Vacua.}
There is of course an important issue that remains to be addressed,
namely whether or not the universe falls into minima of the devaluton field
with values of vacuum energy that are below zero.  Clearly we would not
want to evolve into a universe with a large negative value of the cosmological
constant.  However we show that this does not happen.  Thus the devaluation
process stops near Minkowski space rather than driving the universe to a large
negative vacuum.

We first recall from above that the results of Coleman and de Luccia
\cite{Coleman:1980aw} indicate that the universe will not enter a negative
energy state at the end of the chain inflation period.  Hence the
devaluation stage will be initiated with all regions of space at positive
energy.  While the devaluton may explore regions where its potential
$V_{\rm dev}$ is negative, the thermal energy will always be sufficient to
lift the energy above zero.  Domain walls do not form until the universe
cools.  However, once they do form, they will divide the universe into
separate regions with $\Phi$ in different wells of the potential.  If one
or more of these regions have negative vacuum energy, then they would
presumably end up dominating the universe.  However, there are important
gravitational constraints on the formation of domain walls separating regions
of negative vacuum.  In particular, the radius of curvature of such domain
walls must become ever larger for more negative energy vacua.  Once this
radius of curvature becomes infinite, the domain wall
can no longer form.  Hence this provides a cutoff in how negative the
energy can be on one side of the wall.  Any patches of space with excessive
negative energy cannot become isolated from adjacent regions through domain
wall formation.  Thus they will continue to exchange radiation with the
surrounding regions until they are brought back up to a non-negative
(or only slightly negative) amount of energy.  The result is that it is
impossible for the universe to fall into regions with a cosmological
constant more negative than a critical value.

To understand the cutoff on domain wall formation, we turn again to the
work of Cveti\v{c} {\it et al.}\ \cite{Cvetic1,Cvetic2,Cvetic3} which
provided a careful investigation of domain walls in gravity.  In particular,
these authors found that the vacuum domain walls fall in three classes
depending on their value of wall tension $\sigma$.  These three
classes are determined by whether their wall tension is smaller,
larger, or equal to a value $\sigma_{\rm static}$ determined by
\begin{equation}
\label{eq:sigmaext}
8 \pi G \sigma_{static} = 2 (\alpha_1 \pm \alpha_2),
\end{equation}
where $G$ is Newton's constant, subscripts $1$ and $2$ refer to the two
regions on the two sides of the wall, and the value of the cosmological
constant on either side is $\Lambda_i = - 3 \alpha_i^2$.

The first class, know as extreme walls, occurs when
$\sigma=\sigma_{\rm static}$.  Extreme walls are planar static walls,
corresponding to supersymmetric configurations, and can only interpolate
between Minkowski and Ads (Type I) or AdS and AdS [Type II or III depending on
the sign in (\ref{eq:sigmaext})].  The second class, where
$\sigma > \sigma_{\rm static}$, are know as non-extreme walls.  These
correspond to the interesting situation where observers on both sides of
the domain wall see the wall as receding away from the observer. One can
think of this situation as both observers living inside ever-expanding
bubbles.  Finally, the third class is the ultra-extreme case where
$\sigma < \sigma_{\rm static}$.  Here, an observer in the higher energy
region sees a bubble of lower energy sweeping towards her, so that in the
end only the lower energy region remains.

As ought to be evident, devaluation domain walls fall into the
ultra-extreme category where lower energy regions push aside regions
of higher energy.  In this case the wall tension $\sigma$ is related to
the wall curvature $\beta^2$ according to
\begin{equation}
\label{eq:non}
8 \pi G \sigma/2 = ( - \Lambda_1 /3 + \beta^2 )^{1/2}
- (-\Lambda_2/3 + \beta^2)^{1/2} .
\end{equation}
It is perhaps more illuminating to invert this for the wall curvature.
The result is
\begin{equation}
\label{eq:curv}
\beta^2 = \fft{4 \pi^2\sigma^2}{m_{pl}^4} + \fft{\Lambda_1 + \Lambda_2}6
+ \fft{(\Lambda_2 - \Lambda_1)^2 m_{pl}^4}{64 \times 9 \pi^2 \sigma^2}.
\end{equation}
While $\beta^2\ge0$, we note that the limit $\beta^2\to0$ corresponds
to taking the bubble radius $R\sim\beta^{-1}$ to infinity.  Since a
larger radius is of course impossible, we obtain a bound on how
negative the vacuum energy can be. Setting $\beta^2=0$ in
(\ref{eq:curv}) yields a minimum possible value for the average of the
cosmological constants
\begin{equation}
\label{eq:adsmin}
(\Lambda_1 + \Lambda_2)_{\rm min} = - {3\over 2} \left[ {16 \pi^2 \sigma^2 
\over m_{pl}^4} + {m_{pl}^4 (\Delta \Lambda)^2 \over 16 \times 9 \pi^2
\sigma^2} \right] ,
\end{equation}
where $\Delta \Lambda = |\Lambda_2 - \Lambda_1|$, and any two
neighboring regions have values of the cosmological constant that
differ by $\Delta \Lambda = 8 \pi \epsilon /m_{pl}^2$.  This indicates
that there is a minimum negative value of the cosmological constant;
anything more negative precludes the formation of the domain wall in
the first place.  For example, by taking $\epsilon=(10^{-3}\,\mathrm{eV})^4$
(on the order of the present day cosmological constant), we obtain
\begin{eqnarray}
\label{eq:numbers}
(\rho_{v1} + \rho_{v2})_{\rm min}&=& - (10^{-3}\,\mathrm{eV})^4
\left(\fft\sigma{(10\,\mathrm{MeV})^3}\right)^{-2}\nonumber\\
&&\qquad\times\left(\fft\epsilon{(10^{-3}\,\mathrm{eV})^4}\right)^2,
\end{eqnarray}
where $\rho_{vi} = \Lambda_i/8 \pi G$.
The universe is then left with a small cosmological constant
comparable in magnitude to the present day dark energy density.
The bound here has negative energy density while the dark energy has
a positive value.

Cveti\v{c} {\it et al.}\ have shown that Eqs.~(\ref{eq:sigmaext}) and
(\ref{eq:non}) can also be derived by simple energy arguments. They
define a gravitational (Tolman) mass density $\rho_G = \rho + 3p$.
For a negative vacuum $\rho_G = + 2 |\rho_{vac}|$, {\it i.e.},
attractive. Hence, in the competition between pressure and tension
describing the dynamics of the domain walls, the AdS side exhibits an
attractive ``force'' on the wall. This additional attraction tends to
make small AdS regions disappear \footnote{The opposite behavior is
seen for deSitter bubbles, which have $\rho_G < 0$.  The larger the
positive vacuum, the smaller the bubble radius needs to be in order
for a stable (or dominating) wall to exist.}.  For more negative
AdS, an ever larger radius of curvature is required in order for the
AdS region to be stable or to push aside its higher energy neighbors.
Above we have computed the case of infinite radius to find the most
negative allowed AdS region.

We should note, however, that the results of
Refs.~\cite{Cvetic1,Cvetic2,Cvetic3} depend on simplifying assumptions
that may not hold here.  Firstly, they considered asymptotically dS,
Minkowski and AdS spaces, while our patches may be quite small.  In
addition, the universe in our model has a radiation component that was
not included in their studies.  A complete treatment taking into
account these additional issues would be an important topic for future
study. We also note that we have not considered the case of $\beta^2
<0$, which corresponds to imaginary radius for the domain wall. By
analogy with the work of \cite{Coleman:1980aw}, this case may lead to
singularities and has not yet been examined.

At least up to these caveats, we find that regions of the universe with
large negative cosmological constant are never allowed to come into
existence.  When different patches of the universe fall into different
minima of the potential, they are constrained to fall into minima with
cosmological constant no more negative than allowed by
Eqs. (\ref{eq:adsmin}) or (\ref{eq:numbers}).  Domain wall evolution
then proceeds naturally to lower the cosmological constant to zero (or
just below zero).

{\it Other Issues.}  One might yet worry that some horizon-sized
regions of the universe would fall into very high energy minima of the
devaluton potential and subsequently get trapped without devaluating
down to Minkowski space.  If they inflated, these de Sitter dominated
regions could end up eating up a large portion of the universe, so
that the average spot in the universe would remain in a very high
value of vacuum energy, much higher than is observed.  However, it can
be demonstrated that these potentially dangerous patches do not in
fact cause any problems. Within each horizon sized region (which grows
during radiation domination), there are likely to be many regions, one
of which has a low energy vacuum that takes over the higher energy
ones (as the domain walls sweep them away). In addition, the
population of high energy vacua is Boltzmann suppressed: Larsson {\it
et al} \cite{larsson} have shown that a tiny difference in population
(one part in $10^8$) is sufficient to dynamically drive a two-state
system to the lower energy vacuum.  As these high energy regions are
driven to lower energies, we do not need to resort to anthropic
arguments to explain why we are not in such a region \footnote{For
completeness, we mention that any remaining high energy regions (which
we expect to comprise a minute fraction of the universe) from the
exterior point of view look like black holes or wormholes
\cite{Berezin, Kodama, bgg, fgg,Aguirre:2005xs}; the former may even
contribute to dark matter.}.

The model presented here shares similar features [including the
potential (\ref{eq:pot})] with an earlier proposal by Abbott
\cite{Abbott:1984qf} for dynamically reducing the vacuum energy.  In
his model, the universe begins at a large value of the vacuum energy
and then descends down the potential with decreasing vacuum energy. At
low enough vacuum energy the barriers between minima become important
and the field must tunnel from one minimum to the next lower one.  The
tunneling rate has been computed in the presence of gravity by Coleman
and de Luccia \cite{Coleman:1980aw}, and as mentioned above has been
shown to become slower as the field approaches AdS.  Abbott chose the
parameters of his potential so that the universe ends up with a small
positive cosmological constant, as the lifetime for decay to negative
cosmological constant is longer than the age of the universe. However,
Abbott's model fails in that it gives an empty universe with no
radiation or matter content.  Attempts to revive this model include
\cite{Feng:2000if}, but the problem of an empty universe persists.
Other variants include \cite{brownteit} and \cite{boussopol}. The key
similarity between Abbott's model and ours is that it relies upon the
ability of gravity to differentiate between dS and AdS vacua.  Gravity
knows when the system has reached zero energy and makes it hard to go
lower.

Although our devaluton potential is the same as the one used by
Abbott, the major differences between Abbott's model and ours are as
follows.  First, we have no tunneling between minima of the devaluton
(our second field); instead domain walls sweep aside higher positive
energy vacua in favor of lower energy ones.  Second, we do have
reheating and radiation in our model due to an additional inflaton
field at higher energies.

We briefly discuss what type of particle the devaluton could be.  One
constraint is that it cannot destroy predictions of Big Bang
Nucleosynthesis (BBN). Possibilities include (i) a thermal relic more
massive than $1\,\mathrm{MeV}$, (ii) a particle lighter than $1\,\mathrm{MeV}$
that fell out
of equilibrium before BBN, or (iii) a particle that couples to the
standard model only gravitationally.  In the latter case, the
interactions between the standard and hidden sectors are so weak that
the two sectors decouple at the Planck scale with the two temperatures
then evolving independently.  Such gravitationally coupled devalutons
can be treated as 'shadow matter' as in \cite{kst, enq, ket}.
Candidates for the devaluton also exist in string theory such as (i)
axions and (ii) moduli, which naturally arise in string compactifications
and which may have instanton generated corrections to their potentials.

Domain walls produce temperature anisotropies in the microwave
background of magnitude $\delta T/T \sim G \sigma L$ \cite{zko} where
$L$ is the distance between walls.  Since the walls in our model
disappear via devaluation, these anisotropies are harmless.  The
largest contribution is from horizon sized walls with $L \sim t_d$
where $t_d$ is the universe age at the time the walls disappear,
roughly at $T_d \sim V_0^{1/4}$.  Taking $\delta T/T \leq 10^{-5}$
from observations, we find $\sigma \leq (1\,\mathrm{MeV})^3
(T_d/(10^{-3}\,\mathrm{eV}))^{3/2}$, which may constrain our model.
Structure formation with unstable domain walls has been considered by
\cite{zlor, clo}.

We briefly mention another possibility for bringing the cosmological
constant to a small value.  One could have two tunneling fields which
both chain inflate [{\it e.g.}\ potentials like those of
Eq.~(\ref{eq:chain})], in lieu of an inflaton and a devaluton.  If one
pursues to the extreme the argument that there can be no tunneling to
an AdS vacuum \cite{Banks:2002nm}, one could postulate a tunneling
chain inflaton (leading to reheating) followed by a second chain
tunneling field with tilt $\sim 10^{-3}\,\mathrm{eV}$ that simply stops
tunneling when the potential reaches $V = + 10^{-3}\,\mathrm{eV}$.  Then the
current value of the vacuum energy would be the dark energy and the
cosmological constant problem would be resolved.  However, we find
this solution too simple and have proposed the
devaluation mechanism as more robust and interesting.

{\it Summary.}  We have discussed a dynamical solution to the
cosmological constant problem. After inflation and reheating,
different patches of the universe fall into different nearly
degenerate minima of the devaluton potential separated by domain
walls. We have seen that the domain walls drive the system to the
minimum closest to zero energy.  As long as devaluation begins after
the electroweak scale (the barrier heights of the potential are lower
than the electroweak scale), the potential including all known
contributions to the cosmological constant is driven to a very small
value, in agreement with observations.  The concrete example described
above employed two scalar fields, one for inflation and the other for
devaluation.  However, we note that the basic idea can also work for a
larger number of fields.  For example, there could be a devaluton at
each order of magnitude in mass scale: {\it e.g.} if inflation ends at
the GUT scale, as the temperature drops, first the devaluton with
barrier height 10$^{15}\,\mathrm{GeV}$ becomes important, then the one
with barrier height 10$^{14}\,\mathrm{GeV}$, and so on.  The
interesting thing about this model is that it ties the vacuum energy
of the currently active devaluton to the temperature and may explain
why the dark energy today is near $T_0 \sim 3K$.

Devaluation can easily be generalized within the context of a
multidimensional potential, where the minima of the cosine in this
paper are replaced by vacua (bowls) of the potential separated by
barriers in the multidimensional space.  In an upcoming paper
\cite{abel}, we will construct such a scenario in the context of the
stringy landscape.

\bigskip

We thank S. Abel, A. Aguirre, M. Cveti\v{c}, D. Garfinckel, R. Myers,
L. Pando Zayas, D. Spergel and T. Wiseman for helpful discussions.  We
thank the Michigan Center for Theoretical Physics for hospitality and
support while part of this work was completed.  This work was
supported in part by the US Department of Energy under grant
DE-FG02-95ER40899.


\end{document}